\documentclass[journal]{IEEEtran}
\ifCLASSINFOpdf

\else

\fi

\usepackage{graphicx} %图片
\usepackage{subfigure}
\usepackage{epsfig,epsf,color,balance,cite}

\usepackage[justification=centering]{caption}%浮动体标题
\usepackage{amsmath} %公式
\usepackage{mathtools}
\usepackage{amsfonts} %扩展符号的基础字体支持
\usepackage{amssymb}
\usepackage{bm}%数学符号加粗

\usepackage{algorithmic}
\usepackage{algorithm}%算法
\usepackage{multicol}
\usepackage{booktabs}%排版三线表
\usepackage{tabularx}%tabularx环境排版定宽表格
\usepackage{array,color} %表格列格式扩展

\usepackage{url}
\usepackage{setspace}

\usepackage{comment}   
\usepackage{hyperref}
\usepackage{cite} %参考文献
\usepackage{flushend}  %双栏尾页对称
\usepackage{ntheorem}%定制定理环境
\newtheorem{theorem}{\bf Theorem}
\newtheorem{lemma}{Lemma}
\theorembodyfont{\normalfont}
\newtheorem*{proof}{Proof}

\allowdisplaybreaks[4]  %自动换页

\begin{document}


\title{Reconfigurable Intelligent Surface aided Massive MIMO Systems with Low-Resolution DACs}


%\author{\IEEEauthorblockN{Cunhua Pan, Wence Zhang, Nuo Huang, Houyu Wang and Ming Chen}
%\IEEEauthorblockA{National Mobile Communications Research Laboratory, Southeast University, Nanjing 210096, China\\
%Email:\left\{ cunhuapan, wencehznag, huangnuo, houyuwang and chenming\right\}@seu.edu.cn}
%}

\author{Jianxin Dai, Yuanyuan Wang, Cunhua Pan, Kangda Zhi, Hong Ren, Kezhi Wang
%	\thanks{This work was supported by National 863 High Technology Development Project (No. 2014AA01A701), Key Special Project of National Science and Technology (No.2013ZX03003006), National Natural Science Foundation of China (Nos. 61172077 \& 61372106).}
    \thanks{\textit{(Corresponding author: Cunhua Pan)}.}
	\thanks{J. Dai is with School of Science, Nanjing University of Posts and Telecommunications, Nanjing 210096, China. (email:daijx@njupt.edu.cn).}
	\thanks{Y. Wang is with College of Telecommunications and Information Engineering, Nanjing University of Posts and Telecommunications, Nanjing 210096, China. (email:1219012316@njupt.edu.cn).}
	\thanks{ C. Pan and K. Zhi are with the School of Electronic Engineering and Computer Science at Queen Mary University of London,
	London E1 4NS, U.K. (e-mail: c.pan, k.zhi@qmul.ac.uk).}	
	\thanks{H. Ren is with the National Mobile Communications Research Laboratory, Southeast University, Nanjing 210096, China. 
	(hren@seu.edu.cn).} 
	\thanks{K. Wang is with Department of Computer and Information Science, Northumbria University, UK. 
	(e-mail: kezhi.Wang@northumbria.ac.uk). }
}

\maketitle
%\vspace{-1cm}
\begin{abstract}
We investigate a reconfigurable intelligent surface (RIS)-aided  multi-user massive multiple-input multi-output (MIMO) system where low-resolution digital-analog converters (DACs) are configured at the base station (BS) in order to reduce the cost and power consumption. An approximate analytical expression for the downlink achievable rate is derived based on  maximum ratio transmission (MRT) and additive quantization noise model (AQNM), and the rate maximization problem  is solved by particle swarm optimization (PSO) method under both continuous phase shifts (CPSs) and discrete phase shifts (DPSs) at the RIS. Simulation results show that the downlink sum achievable rate tends to a constant with the increase of the number of quantization bits of DACs, and four quantization bits are enough to capture a large portion of the performance of the ideal perfect DACs case.

% In addition, the performance loss caused by low-resolution DACs can be compensated by increasing the number of transmit antennas.
\end{abstract}

%\begin{keywords}
% Energy efficiency, adaptive price, distributed algorithm, interference channel.
%\end{keywords}


% \begin{center} \pmbseries EDICS Category: 3-BBND \end{center}
% \fi
%\IEEEpeerreviewmaketitle
%\newpage
%\baselineskip=8.8mm

\vspace{-0.3cm}\section{Introduction}

Recently, a reconfigurable intelligent surface (RIS), which can configure the wireless propagation environment, has attracted extensive research interests \cite{Huang_Holographic,MD2019Smart,9366805}.
Specifically, the RIS is a planar consisting of a large number of passive elements, each of which can manipulate the electromagnetic characteristics of reflected signal independently. By carefully tuning the phase shifts of the RIS, the reflected signals can be constructively added with the direct signals from the BS to enhance the desired signal power, or destructively added with the direct signal to mitigate the undesired signals.
%By carefully tuning the phase shifts of the RIS, the reflected signals can be constructively added with the direct signals from the BS to enhance the desired signal power, or destructively added with the direct signal to mitigate the undesired signals. 
Some advantages of RIS-assisted wireless communications include: easy deployment, low hardware cost, enhanced energy- or spectrum-efficiency (EE/SE), and easy integration into the existing networks \cite{Qtowards}.  As a result, the RIS is expected to push forward an immense influence on improving the transmission performance of future wireless communication networks.

To reap the benefits promised by the RIS, the phase shifts of the reflecting elements at the RIS should be carefully designed. Most of the existing contributions designed the phase shifts based on the instantaneous channel state information (CSI) \cite{Qintelligent,panmulticell,panintelligent}. This scheme has some drawbacks. Firstly, the phase shifts of the RIS reflecting elements need to be calculated within each channel coherence time that varies rapidly (on the order of milliseconds). This will incur high computational complexity at the base station (BS). Secondly, this scheme requires to estimate instantaneous cascaded CSI that will entail high channel estimation overhead, the amount of which generally increases linearly with the number of reflecting elements. Hence, it is unaffordable for the scenario when the channel coherence time is very short. Thirdly, since the phase shifts need to be updated for each channel coherence time, there will be frequent information exchange between the BS and the RIS. One promising solution to addressing these drawbacks is to design the phase shifts based on the long-term CSI such as angle information \cite{yuhanlarge,9366346,zhi2021power,Huang9110869} or location information \cite{Xhu2020Location}, which varies much more slowly than instantaneous CSI.

All the above-mentioned contributions  \cite{yuhanlarge,9366346,zhi2021power,Huang9110869, Xhu2020Location} assumed the ideally perfect hardware at the BS. 
Authors in \cite{2019Reconfigurable} investigated the energy efficiency maximization problem for RIS-aided multi-user communication where the static power of hardware was taken into account.
In massive MIMO systems, each antenna is connected to one analog-to-digital converter (ADC) or digital-to-analog converter (DAC), and will consume high power consumption when adopting high-resolution ADC/DACs due to the large number of antennas. Hence, it is appealing to adopt low-resolution ADC/DACs in massive MIMO systems due to its reduced cost and low power consumption \cite{fan2015uplink}. 

%Most recently, the authors in \cite{kangdauplink} derived the uplink achievable rate expression of RIS-aided millimeter-wave (mmWave) systems by taking into consideration the quantization error at the BS and the phase noise at the RIS.

Against the above background, in this paper we study a  RIS-aided multi-user massive MIMO system, where the BS is equipped with low-resolution DACs. Specifically, our contributions are summarized as follows:

\begin{enumerate}
  \item We derive downlink sum achievable rate of the multi-user massive MIMO system based on Rician channel model;
  \item We utilize particle swarm optimization (PSO) algorithm to solve the achievable rate maximization problem by optimizing the phase shifts by considering both continuous phase shifts (CPSs) and discrete phase shifts (DPSs);
  \item Through simulations, we analyze the impacts of the number of quantization bits of DACs and phase shifts at the RIS on the rate performance, and verify the effectiveness of the proposed algorithm.
\end{enumerate}

%The rest of the paper is organized as follows.
%In Section \uppercase\expandafter{\romannumeral2}, we introduce the RIS-aided multi-user massive MIMO communication model.
%We derive the downlink achievable rate in Section  \uppercase\expandafter{\romannumeral3}  and optimize the phase shifts in Section  \uppercase\expandafter{\romannumeral4}. The simulation results are given and analyzed in Section  \uppercase\expandafter{\romannumeral5}. Conclusions are drawn in Section  \uppercase\expandafter{\romannumeral6}.

\textbf{Notations:}   
%For matrix $\mathbf{X}$ and vector $\mathbf{x}$,   $[\mathbf{X}]_{mn}$ and $[\mathbf{x}]_n$ denote the ($ m,n $)-th entry of the matrix and the $ n $-th entry of the vector, respectively.
\textrm{diag}$(\mathbf{x})$ denotes a diagonal matrix with the entries of $\mathbf{x}$ on its main diagonal. The symbols $\mathbb{E} \{ \cdot \}$,  $\textrm{Re}\{ \cdot \}$, and $ \textrm{Tr}( \cdot )$ denote  the expectation operator, real part, and trace, respectively. $\mathbf{I}_N$ is the identity matrix with dimension of $ N $ . $ \mathbb{C}^{M \times N} $ represents the $ M \times N $ complex-valued
matrix. Besides, $x \sim \mathcal{CN} (a,b)$ denotes that random variable $ x $ follows the complex Gaussian
distribution with mean $ a $ and variance $ b $. 
%$ j \triangleq \sqrt{-1} $ is the  imaginary unit.

%Specifically, an RIS architecture is proposed in [3] to achieve amplitude-and-phase-varying modulation, which facilitates the design of multiple-input multiple-output (MIMO) quadrature amplitude modulation (QAM) transmission.

\section{System Model}\label{model}
\vspace{-0.3cm}
{\setlength{\abovecaptionskip}{-0.2cm} 
	\setlength{\belowcaptionskip}{-0.3cm} 
\begin{figure}[H]
	\centering
	\includegraphics[scale=0.7]{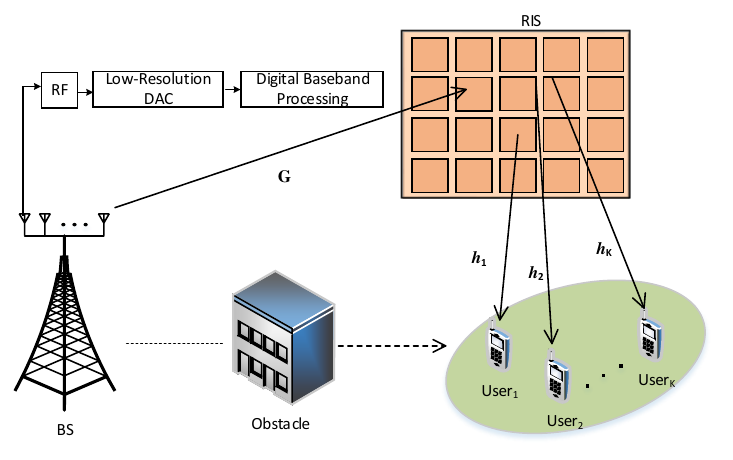}
	\caption{System Model}
	\label{System Model}
\end{figure}}
We consider a downlink multi-user massive MIMO system where a passive RIS with discrete unit elements is deployed to assist the communication from an $M$-antenna BS to $K$ single-antenna mobile users, as shown in Fig. \ref{System Model}. %\cite{Huang_Holographic}.
The BS consists of a large-scale uniform linear antenna array (ULA) and each antenna is equipped with a low-resolution DAC, and the RIS is composed of $N$ reflecting elements.
We assume that the direct link between the BS and users is neglected due to obstacles,  
and the RIS is deployed at a proper position where line-of-sight (LoS) communication is ensured for both BS-to-RIS and RIS-to-user links.

%It is assumed that the CSI of all links is perfectly known at the BS.
Let $ \mathbf{G} \in \mathbb{C}^{N \times M} $,
$\mathbf{H} = \left[\mathbf{h}_1,\mathbf{h}_2, \dots ,\mathbf{h}_K \right] \in \mathbb{C}^{N \times K} $ respectively denote the channel matrices for the channel between the BS and the RIS, and that  between the RIS and $ K $ users, and the Rician fading model is adopted for the channel. Here, $ \mathbf{h}_k^H \in \mathbb{C}^{1 \times N} $ represents the channel from the RIS to the $k$-th user. Specifically,
$ \mathbf{G} $ and $ \mathbf{h}_k^H $ can be expressed as:
{\setlength\abovedisplayskip{3pt}	
\setlength\belowdisplayskip{3pt}
\begin{equation}
	\mathbf{G} =\sqrt{\varepsilon}\left( \sqrt{\frac{K_G}{K_G+1}} \bar{\mathbf{G}} + \sqrt{\frac{1}{K_G+1}}  \widetilde{\mathbf{G}}\right)
\end{equation}}
and
{\setlength\abovedisplayskip{3pt}	
\setlength\belowdisplayskip{3pt}
\begin{equation}
	\mathbf{h}_k^H =\sqrt{\beta_k} \left( \sqrt{\frac{K_k}{K_k+1}} {\bf{\bar h}}_k^H + \sqrt{\frac{1}{K_k+1}}  \tilde{\mathbf{h}}_k^H\right),
\end{equation}}
where
$\varepsilon $ and $\beta _k $ represent the distance-dependent path loss of BS-to-RIS and RIS-to-$ k $-th-user paths, and $ K_G $, $ K_k $ refer to Rician factors.
$ \tilde{\mathbf{G}} $, $ \tilde{\mathbf{h}}_k^H $ are scattering components, each element of which is i.i.d. complex Gaussian distributed with zero mean and unit variance and $ \bar{\mathbf{G}} $, $ \bar{\mathbf{h}}_k^H $ are LoS components, which can be expressed by the responses of the ULA as:
{\setlength\abovedisplayskip{3pt}	
\setlength\belowdisplayskip{3pt}
\begin{equation}
	\bar{\mathbf{G}}=  \mathbf{a}_N\left( \phi_r\right) \mathbf{a}_M^H \left( \phi_t\right) ,\qquad \bar{\mathbf{h}}_k^H=  \mathbf{a}_N^H \left( \varphi_{kt}\right),
\end{equation}}
where
$\phi_r $ is the angle of arrival (AoA) at the RIS,  $\phi_t $ and $ \varphi_{kt}  $ are respectively angle of departure (AoD) at the BS and the  $k$-th user's AoD at the RIS.
In addition, the array response of an $ X $-element ULA is:
{\setlength\abovedisplayskip{3pt}	
	\setlength\belowdisplayskip{3pt}
\begin{equation}
	\mathbf{a}_X\left( \vartheta \right) = \left( 1,e^{j2 \pi \frac{d}{ \lambda} \sin \vartheta } , \dots , e^{j2 \pi \frac{d}{ \lambda} \left(X-1\right) \sin \vartheta} \right)^T,
\end{equation}}
where $ d $ and $ \lambda $ are the element spacing and signal wavelength. In this paper, we assume that the  statistical CSI can be readily obtained by the existing channel estimation methods.

Define an  $ N \times N $ diagonal matrix
$ \mathbf{\Phi} = \textrm{diag} \left( \zeta_1 \mathbf{e}^{j \theta _1} , \zeta_2 \mathbf{e}^{j \theta _2}, \dots , \zeta_N \mathbf{e}^{j \theta _N} \right)  $ as the reflection coefficient matrix of the RIS, where
$ \zeta _n \in \left[0,1\right] $ and
$ \theta _n \in [0,2 \pi) $ for $ n=1,2,\dots,N $ represent the amplitude reflection efficiency and the phase shifts induced by the $n$-th reflecting unit, respectively. Without loss of generality,  we set $ \zeta _n =1 $ for all $n$.

The unquantized downlink transmission signal at the BS can be written as:
{\setlength\abovedisplayskip{3pt}	
	\setlength\belowdisplayskip{3pt}
\begin{equation}\label{transmit signal}
	\mathbf{x} = \mathbf{W} \mathbf{s},
\end{equation}}
where
$ \mathbf{s} = \left( s_1 ,s_2 , \dots ,s_K \right) ^T \in \mathbb{C}^{K \times 1} $ denotes the transmit signal vector of the BS, which satisfies
$ \mathbb{E}\left\{ \mathbf{s} \mathbf{s}^H \right\} = \mathbf{I}_K $, and
$ \mathbf{W} = \left[ \mathbf{w}_1 , \mathbf{w}_2 \dots \mathbf{w}_K\right] \in \mathbb{C}^{M \times K} $ denotes the precoding matrix.

Based on additive quantization noise model (AQNM), the downlink transmission signal quantized by DACs at the BS can be expressed as:
{\setlength\abovedisplayskip{3pt}	
	\setlength\belowdisplayskip{3pt}
\begin{equation}\label{quantized signal}
	\mathbf{x}_q = Q\left( \mathbf{x} \right) = \alpha \mathbf{x} + \mathbf{n}_q,
\end{equation}}
where
$ Q \left( \cdot\right) $ is a quantization function \cite{ylidownlink}, with
$ \alpha = 1- \rho $, where
$ \rho $ is the inverse of the signal-to-quantization-noise ratio and
$ \mathbf{n}_q \in \mathbb{C}^{M \times 1} $ denotes the additive Gaussian quantization noise, that is uncorrelated with $ \mathbf{x} $,  whose covariance is:
{\setlength\abovedisplayskip{3pt}	
	\setlength\belowdisplayskip{3pt}
\begin{equation}\label{R_nq}
	 \mathbf{R}_{\mathbf{n}_q\mathbf{n}_q} 
	 = \mathbb{E}\left\{ \mathbf{n}_q\mathbf{n}_q^H \right\} 
	 =\alpha \left(1-\alpha\right) \textrm{diag} \left( \mathbf{W} \mathbf{W} ^H\right).
\end{equation}}
The values of $\rho $ corresponding to the quantization bits $b$ are listed in Table I for $ b \leq 5 $ and can be approximated by
$ \rho = \frac{\sqrt{3} \pi }{2}  \cdot 2^{-2b} $ for $b > 5$ \cite{fan2015uplink}.
\begin{table}[H]
	\setlength{\abovecaptionskip}{0cm} 
	\setlength{\belowcaptionskip}{-0.2cm}
	\caption{ $\rho$ VERSUS QUANTIZATION BITS $ b $}
	\begin{center}
		\setlength{\tabcolsep}{2mm}{
		\begin{tabular}{cccccc}
			\toprule
			$ b $ & 1 & 2 & 3 & 4 & 5   \\
			\midrule
			$\rho $ & 0.3634 & 0.1175 & 0.03454 & 0.009497 & 0.002499\\
			\bottomrule
		\end{tabular}}
		%   \label{tab:my_label}
	\end{center}
\end{table}
\vspace{-0.5cm}
Therefore, the downlink received signal of the users can be expressed as:
{\setlength\abovedisplayskip{3pt}	
	\setlength\belowdisplayskip{3pt}
\begin{equation}
	\mathbf{y} = \sqrt{P} \mathbf{F}^H \mathbf{x}_q + \mathbf{n},
	%\qquad \text{k=1,2,\dots,K }
\end{equation}}
where $ \mathbf{F}^H =  \mathbf{H}^H  \mathbf{\Phi}  \mathbf{G} \in \mathbb{C}^{K \times M} $ represents the cascaded BS-RIS-user channel, 
$ P $ represents the transmit power at the BS, and $ \mathbf{n} \sim  \mathcal{CN}\left( 0,\sigma^2 \mathbf{I}_K \right) $ denotes the AWGN vector at the users.

By using \eqref{transmit signal} and \eqref{quantized signal}, the received signal of the $k$-th user is given by:
{\setlength\abovedisplayskip{3pt}	
	\setlength\belowdisplayskip{3pt}
\begin{equation}\label{received signal}
y_k \!=\!  \alpha  \sqrt{P} \mathbf{f}_k^H \mathbf{w}_k s_k + \alpha  \textstyle \sum_{
	i=1,i \ne k
	}^K \! \sqrt{P} \mathbf{f}_k^H \mathbf{w}_i s_i \!+\! \sqrt{P} \mathbf{f}_k^H \mathbf{n}_q \!+\! n_k,
%\qquad \text{k=1,2,\dots,K }
\end{equation}}
where
$ \mathbf{f}_k^H = \mathbf{h}_k^H   \mathbf{\Phi}  \mathbf{G} $, with $k=1,2,...,K$.
The first term on the right hand side of \eqref{received signal} is the desired signal, the second term is the multi-user interference, the third term is quantization noise and the last term is the AWGN.

\begin{comment}
	\begin{lemma}
	if $ \mathbf{X} = \sum_{i=1}^{t1} X_i $ and $ \mathbf{Y} = \sum_{j=1}^{t2} Y_j $ are  the sum of non-negative random variables $X_i$ and $Y_j$, then we can get the following approximate expression:
	\begin{equation}
	\mathbb{E} \left\{  \log_{2}{ \left( 1 + \frac{ \mathbf{X}}{ \mathbf{Y}} \right)} \right\}
	\approx
	\log_{2}{\left( 1+ \frac{ \mathbb{E} \left\{ \mathbf{X}\right\}  }{\mathbb{E} \left\{ \mathbf{Y}\right\} }\right)}
	\end{equation}
	Note that the approximation here, which will become more accurate as $ t_1 $ and $ t_2 $ increase, does not require random variables $\mathbf{X}$ and $\mathbf{Y}$ to be independent.
	\end{lemma}
	??...
\end{comment}

\vspace{-0.3cm}\section{Analysis of Achievable Rate}\label{analysis}

In this paper, the MRT method is adopted to process the transmit signal at the BS to maximize the signal power gains of the desired users. Then, the precoding matrix $\mathbf{W}$ of the BS  is:
{\setlength\abovedisplayskip{3pt}	
	\setlength\belowdisplayskip{3pt}
\begin{equation}
	\mathbf{W}= \frac{ \mathbf{F} }{ \sqrt{\textrm{Tr} \left( \mathbf{F}^H \mathbf{F} \right)} }.
\end{equation}}

From \eqref{received signal}, we can obtain the signal-to-interference-plus-noise ratio (SINR) at the $k$-th user, which can be expressed as:
{\setlength\abovedisplayskip{1pt}	
	\setlength\belowdisplayskip{3pt}
\begin{equation}
	\begin{aligned}
		\gamma_k ={}& \frac{ \alpha^2 P \left|\mathbf{f}_k^H \mathbf{w}_k  \right|^2 }
		{ \alpha^2 P \sum_{i=1,i \neq k}^K  \left|\mathbf{f}_k^H  \mathbf{f}_i \right|^2 + P \left| \mathbf{f}_k^H \mathbf{n}_q \right|^2 +  |n_k|^2  }\\
%		={}& \frac{ \alpha^2 P \left|\mathbf{f}_k^H \mathbf{f}_k \mathbf{s}_k/\sqrt{\textrm{Tr} \left( \mathbf{F}^H \mathbf{F} \right)} \right|^2 }
%		{ \alpha^2 P \sum_{i=1,i \neq k}^K  \left|\mathbf{f}_k^H \mathbf{f}_i \mathbf{s}_i/\sqrt{\textrm{Tr} \left( \mathbf{F}^H \mathbf{F} \right)} \right|^2 + P | \mathbf{f}_k^H \mathbf{n}_q|^2 +  |n_k|^2  }  \\
		={}& \frac{ \alpha^2 P \left \|\mathbf{f}_k \right\|^4 }
		{ \alpha^2 P \sum_{i=1,i \neq k}^K  \left|\mathbf{f}_k^H \mathbf{f}_i  \right|^2 + P \Gamma  | \mathbf{f}_k^H \mathbf{n}_q|^2  + \Gamma |n_k|^2} ,
	\end{aligned}
\end{equation}}
where  $ \Gamma= \textrm{Tr} \left( \mathbf{F}^H \mathbf{F} \right) $. Therefore, the achievable rate of the $k$-th user can be expressed as:
{\setlength\abovedisplayskip{3pt}	
	\setlength\belowdisplayskip{3pt}
\begin{equation}\label{R_k}
R_k = \mathbb{E}\left\{ \log_{2}{\left(1+\gamma_k\right)} \right\}.
\end{equation}}
A closed-form approximation of \eqref{R_k} is obtained in the following theorem and the sum rate can be written as:
{\setlength\abovedisplayskip{3pt}	
	\setlength\belowdisplayskip{3pt}
\begin{equation}
R_{\rm{sum}} = \sum_{k=1}^K R_k.
\end{equation}}
\vspace{-0.4cm}
\begin{theorem}
In the RIS-aided massive MIMO System with Low-Resolution DACs, the downlink achievable rate can be approximated as:
%{\setlength\abovedisplayskip{3pt}	
%	\setlength\belowdisplayskip{3pt}
\vspace{-0.2cm}
\begin{equation}\label{R_k_closed}
	R_k \approx  \log_{2}{\left( 1+\frac{ \alpha^2 P E_k^{signal} }{ \alpha^2 P \sum_{i = 1, i \neq k}^K I_{ki} + P I_k^{DAC} + \sigma^2   E_k^{noise}}\right) },
\end{equation}
where
$ E_k^{signal} $, 
$ I_{ki}  $, 
$ I_k^{DAC} $, and
$ E_k^{noise} $
are respectively given by \eqref{signal}, \eqref{interference}, \eqref{DAC_interference} and \eqref{noise}. Besides, 
$ \delta_c =\frac{\varepsilon \beta_c  }{\left(K_G+1\right)\left(K_c+1\right)} $,  
$ \psi_c(\mathbf{\Phi})=\mathbf{a}_N^H\left( \phi_r\right)\mathbf{\Phi}^H \bar{\mathbf{h}}_c
= \sum_{n=1}^{N} e^{j 2\pi \frac{d}{\lambda}(n-1)(\sin(\varphi_{ct})-\sin(\phi_r)  ) - j\theta_n  }  
\in \mathbb{C}^{1 \times 1} $, 
$c \in \{k,i\} $. 
\end{theorem}
\vspace{-0.3cm}
{\setlength\abovedisplayskip{3pt}	
	\setlength\belowdisplayskip{3pt}
\begin{figure*}[!t]
	{\setlength\abovedisplayskip{3pt}	
		\setlength\belowdisplayskip{3pt}
		\vspace{-0.3cm}
	\begin{equation}\label{signal}	
	\begin{aligned}
		&E_k^{signal} 
		= M \delta_k^2 \times 
		\Big\{  {M (K_G K_k)^2 | \psi_k(\mathbf{\Phi})|^4} 
		+ {2K_GK_k | \psi_k(\mathbf{\Phi})|^2 ( 2MNK_G + MNK_k + MN + 2M+ NK_k + N + 2 )}    \\
		&+ {MN^2( 2K_G^2 + K_k^2 +2K_G K_k + 2K_G + 2K_k +1 )} 
		+ {N^2( K_G^2 + 2K_GK_k + 2K_G +2K_k+1 )} \\
		&+ MN( 2K_G+ 2K_k +1) + N(2K_G+2K_k+1) 
		\Big\}, 
	\end{aligned}
\end{equation}}
\vspace{-0.1cm}
\begin{equation}\label{interference}
	\begin{aligned}	
		&I_{ki}
		= M \delta_k \delta_i \times 
		\Big\{ 
		{M K_G^2 K_k K_i |\psi_k(\mathbf{\Phi})|^2 | \psi_i(\mathbf{\Phi})|^2 }
		+ {K_G K_k | \psi_k(\mathbf{\Phi})|^2 (K_GMN + NK_i +N +2M)} \\
    	&+ {K_G K_i | \psi_k(\mathbf{\Phi})|^2 (K_GMN + NK_k +N +2M)}   + {N^2 ( MK_G^2 + K_G(K_k+K_i+2) + (K_k+1)(K_i+1))} \\ 
		&+ {MN(2K_G+K_k+K_i+1)} 
		+ {MK_kK_i \left| \bar{\mathbf{h}}_k^H \bar{\mathbf{h}}_i^H \right|^2} 
		+ {2MK_GK_kK_i \textrm{Re}\left\{ \psi_k^H (\mathbf{\Phi}) \psi_i (\mathbf{\Phi}) \bar{\mathbf{h}}_i^H \bar{\mathbf{h}}_k\right \}	}	
		\Big\},
	\end{aligned}
\end{equation}
\vspace{-0.1cm}
{\setlength\abovedisplayskip{3pt}	
	\setlength\belowdisplayskip{3pt}
\begin{equation}\label{DAC_interference}
	\begin{aligned}
		&I_k^{DAC}=
		 \alpha \left( 1-\alpha \right) M \times 
		\Big\{ 
		\delta^2 
		\Big\{
		\left( K_G K_k \left| \psi_k(\mathbf{\Phi}) \right|^2 \right)^2 		
		\!+\! 2N^2 \left(K_G \!+\!K_k \!+\!1 \right)^2 
		\!+\! 4K_G K_k  \left| \psi_k(\mathbf{\Phi}) \right|^2 \left( N (K_G \!+\!K_k\!+\!1)+2 \right)  \\	
		&\!+\! 2N(2K_G+2K_k+1) \Big\} 
		+ \textstyle\sum_{i=1,i \ne k }^K \delta_k \delta_i \Big\{{ K_G^2 K_k K_i| \psi_k(\mathbf{\Phi})|^2 | \psi_i(\mathbf{\Phi})|^2 } 
		+ {N K_G K_k |\psi_k(\mathbf{\Phi})|^2(K_G + K_i +1)} \\
		&+ {N K_G K_i |\psi_i(\mathbf{\Phi})|^2(K_G + K_k +1)} 
		+ {N^2( K_G^2 +K_G K_k + K_G K_i +K_k K_i +2K_G +K_k +K_i +1 )}
		\Big\}
		\Big\}, 
	\end{aligned}	
\end{equation}}
and
\vspace{-0.2cm}
{\setlength\abovedisplayskip{2pt}	
	\setlength\belowdisplayskip{2pt}
\begin{equation}\label{noise}
	\begin{aligned}	
		E_k^{noise}
		= M \textstyle \sum_{k=1}^{K} \delta_k \left( K_G K_k  | \psi_k(\mathbf{\Phi})|^2 + N \left( K_G + K_k +1 \right)  \right).
	\end{aligned}
\vspace{-0.3cm}
\end{equation}}
\vspace{-0.3cm}
\hrulefill
\end{figure*}}
\vspace{-0.1cm}
{\setlength\abovedisplayskip{3pt}	
	\setlength\belowdisplayskip{3pt}
\begin{proof}
By applying \cite[Lemma 1]{qzhangpower} and \eqref{R_nq}, the achievable rate can be approximated as \eqref{R_k_appro}.
	{\setlength\abovedisplayskip{3pt}	
	\setlength\belowdisplayskip{3pt}
\begin{figure*}
\begin{equation}\label{R_k_appro}
	R_k \approx  \log_{2}{\left( 1+  
	\frac{ {\alpha^2 P  \mathbb{E} \left\{ \| \mathbf{f}_k \|^4 \right\} }   }{ {\alpha^2 P \sum_{i = 1, i \neq k}^K \mathbb{E} \left\{ | \mathbf{f}_k^H \mathbf{f}_i |^2 \right\}  } 
	+ {P \mathbb{E} \left\{ \Gamma  \mathbf{f}_k^H \mathbf{R}_{\mathbf{n}_q\mathbf{n}_q} \mathbf{f}_k  \right\}} 
	+  \sigma^2 \mathbb{E} \left\{ \Gamma \right\}     } \right) }.
\end{equation}
\vspace{-0.6cm}
\end{figure*}}
 
To derive the closed-form expression, we need to derive signal term $ \mathbb{E} \left\{ \| \mathbf{f}_k \|^4 \right\} $, interference term $ \mathbb{E} \left\{ | \mathbf{f}_k^H \mathbf{f}_i |^2 \right\} $, quantization noise  term $ \mathbb{E} \left\{ \Gamma \mathbf{f}_k^H
\mathbf{R}_{\mathbf{n}_q\mathbf{n}_q} \mathbf{f}_k \right\} $ and AWGN noise term $ \mathbb{E} \left\{ \Gamma \right\}   $ after simplification. 
Define $ \mathbf{f}_{km} $ as the $ m $-th entry of $ \mathbf{f}_k $, the first two terms have been given in \cite[Lemma 1]{zhi2021power} and the remaining two terms are derived as : 
{\setlength\abovedisplayskip{4pt}	
	\setlength\belowdisplayskip{1pt}
\begin{align}\label{E_qua noise}
%	\begin{aligned}
		% \mathbb{E}  \left\{ T | \mathbf{f}_k^H \mathbf{n}_q|^2  \right\} 
		 &\mathbb{E} \left\{ \Gamma \mathbf{f}_k^H
		\mathbf{R}_{\mathbf{n}_q\mathbf{n}_q} \mathbf{f}_k \right\}   
		 = \alpha \left( 1-\alpha \right) \mathbb{E} \left\{  \mathbf{f}_k^H  \textrm{diag} \left( \mathbf{F} \mathbf{F}^H   \right)  \mathbf{f}_k  \right\}  \notag \\
		 &= \alpha \left( 1-\alpha \right) \mathbb{E} \Big\{\! \textstyle \sum_{m=1}^{M} \! \left| \mathbf{f}_{km}  \right|^2 \!\Big(\! \left|\mathbf{f}_{km} \right|^2 \!+\!  \sum_{i=1,i \ne k }^K \! \left|\mathbf{f}_{im} \right|^2   \!\Big)\!  \Big\}   \notag\\ 
		&= \alpha \left( 1-\alpha \right) \Big(\textstyle \sum_{m=1}^{M} \mathbb{E} \big\{ \left| \mathbf{f}_{km} \right|^4 \big\}  
		+ \sum_{m=1}^{M} \sum_{i=1,i \ne k }^K  \notag\\ 
		&\mathbb{E}\big\{ \left| \mathbf{f}_{km} \right|^2  \big\} \mathbb{E}\big\{ \left| \mathbf{f}_{im} \right|^2  \big\} \Big),  
%	\end{aligned}	
\end{align}}
\vspace{-0.1cm}
where 
% E{|f_km|^2}
{\setlength\abovedisplayskip{3pt}	
	\setlength\belowdisplayskip{2pt}
\begin{equation}\label{E_|f_km|^2}
	\begin{aligned}
		&\mathbb{E} \big\{ \left| \mathbf{f}_{km} \right|^2 \big\} 
		= \frac{1}{M} \mathbb{E} \left\{ \| \mathbf{f}_k \|^2 \right\} \\
	&	= \delta_k \left( K_G K_k  | \psi_k(\mathbf{\Phi})|^2 \!+\! N \left( K_G \!+\! K_k \!+\!1 \right)  \right),
	\end{aligned}
\end{equation}}
% E{|f_km|^4}
{\setlength\abovedisplayskip{3pt}	
	\setlength\belowdisplayskip{3pt}
\begin{align}\label{E_|f_km|^4}
%	\begin{aligned}
	&\mathbb{E} \big\{ \left| \mathbf{f}_{km} \right|^4 \big\} 
	\!=\! \delta^2 
	\Big\{
	\left( K_G K_k \left| \psi_k(\mathbf{\Phi}) \right|^2 \right)^2 		
	\!+\! 2N^2 \left(K_G \!+\!K_k \!+\!1 \right)^2 \notag \\\vspace{-0.4cm}
	&\!+\! 4K_G K_k  \left| \psi_k(\mathbf{\Phi}) \right|^2 \left( N (K_G \!+\!K_k\!+\!1)+2 \right)  \\	
	&\!+\! 2N(2K_G+2K_k+1) \Big\}, \notag
%	\end{aligned}
\end{align}}
which can be derived by applying \cite[Lemma 1]{zhi2021power}. In addition, we have
{\setlength\abovedisplayskip{3pt}	
	\setlength\belowdisplayskip{2pt}
%\begin{equation}\label{E_noise}		
	\begin{align}\label{E_noise}	
		&\mathbb{E} \left\{ \Gamma \right\}
		\!=\!  \mathbb{E}\! \Big\{ \textstyle\sum_{k=1}^{K} \sum_{m=1}^{M} \left|  \mathbf{f}_{km}  \right|^2 \! \Big\}  
		\!=\! \sum_{k=1}^{K} \sum_{m=1}^{M} \mathbb{E}\! \left\{ \left| \mathbf{f}_{km}  \right|^2  \right\} \notag \\ 
		&\!=\! M \textstyle\sum_{k=1}^{K} \delta_k \left( K_G K_k  | \psi_k(\mathbf{\Phi})|^2 \!+\! N \left( K_G \!+\! K_k \!+\! 1 \right)  \right).
	\end{align}
%\end{equation}
}
By substituting \eqref{E_qua noise}, \eqref{E_noise} and the useful signal term and interference term into \eqref{R_k_appro}, we can obtain the final result. This completes the proof.
\end{proof}}

\vspace{-0.3cm}\section{Phase Shift Optimization}\label{phaseshiftopti}
 
%Both CPSs and DPSs constraints are considered in this paper, and the results obtained from the continuous phase shift can serve as the performance benchmark for the practical case of discrete phase shift.  We aim to maximize the sum achievable rate by optimizing the phase shifts based on the long-term CSI, which can be formulated as: 
 
In this section, we aim to maximize the sum achievable rate by optimizing the phase shifts, considering both CPSs and DPSs, based on the long-term CSI, which can be formulated as:
\vspace{-0.1cm}
\begin{equation}\label{optimization}
	\begin{aligned}
		&\max \limits_{\mathbf{\Phi}}\quad R_{\rm{sum}}\\
		& \begin{array}{r@{\quad}r@{}l@{\quad}l}
			\rm{s.t.}& \theta _n \in {\cal F}_1\  \rm{or} \ {\cal F}_2, \forall n = 1,2, . . . , N.\\
		\end{array}
	\end{aligned}
\end{equation}
where ${{\cal F}_1} \!=\! \left\{ {\left. {{\theta _n}} \right|0 \le {\theta _n} \le 2\pi } \right\}$ and  ${{\cal F}_2}\! = \!\big\{ {0,\frac{{2\pi }}{{{2^B}}}, \cdots ,\frac{{2\pi \left( {{2^B} - 1} \right)}}{{{2^B}}}} \big\}$ denote the sets of continuous and discrete phase shift values, respectively. Here, we limit the period phase in order to simplify the algorithm and $B$ denotes the number of quantization bits of phase shifts at the RIS. 
%In the following, we first obtain the solution under the continuous case, then the discrete values can be obtained by directly quantizing the continuous solutions to its nearest discrete values.

Due to the complex expression of the objective function, 
%it is challenging to directly optimize the phase shift matrix by using conventional optimization methods. To address this problem, 
we adopt a PSO algorithm in Algorithm \ref{alg:PSO} owing to its high universality. 
Supposing the size of particle population is $ L$, and the maximum number of iterations is $ T $, the complexity of the algorithm is proportional to $ L \times T $ \cite{2014Low}.  
%the complexity of the algorithm is proportional to $ N \times T $.
For $ i=1,2,\dots,L $, the coordinate position of particle $ i $ at time $ t $ can be associated to a $ 1 \times N $ phase shift vector $  {\bm{\theta }}_i^{(t)} = (\theta_{i1}, \theta_{i2}, \dots , \theta_{iN} ) $, each element of which is generated randomly limited within $ {\cal F}_1 $ or $ {\cal F}_2 $.
Specifically, the difference between $ {{\cal F}_1} $ and $ {{\cal F}_2} $ is whether the phase is discretized during initialization.

{\setlength{\abovecaptionskip}{-0cm} 
	\setlength{\belowcaptionskip}{-0.2cm}  
	\begin{algorithm}[t]
		\caption{Particle Swarm Optimization Algorithm (PSO)}
		\label{alg:PSO}
		\setstretch{1}
		\begin{algorithmic}[1]
			\setlength{\abovecaptionskip}{-0.2cm} 
			\setlength{\belowcaptionskip}{-0.2cm}
			\STATE Particle swarm parameters initialization: 
			Initialize $ L $, $ T $, $ \omega$, $ c_1$, $c_2$, $r_1$ ,$r_2 $;\\
			
			\FOR{$ i=1,2,\dots,L  $}
			\STATE Initialize  
			$ \bm{\theta}_i^{(0)} $, $ \mathbf{v}_i^{(0)} $,
			$ \mathbf{p}_i^{(0)} = \bm{\theta}_i^{(0)} $;  	\\
			\ENDFOR
			
			\STATE Find $R'( \mathbf{p}^{*(0)} ) = \min\{ R'( \mathbf{p}_1^{(0)}),\dots,R'( \mathbf{p}_L^{(0)}) \}    $, and set $ \mathbf{g^{(0)}}= \mathbf{p}^{*(0)}$;\\
			
			\WHILE{$ t \le T $}
			\FOR{ $ i=1,2,\dots,L $}
			
			\STATE Update the velocity and position of particles: \\
			$ \mathbf{v}_i^{(t+1)} =  \omega \mathbf{v}_i^{(t)} + c_1 r_1( \mathbf{p}_i - \bm{\theta}_i^{(t)} ) +  c_2 r_2( \mathbf{g} - \bm{\theta}_i^{(t)})  $, \\
			$ \bm{\theta}_i^{(t+1)} =  \bm{\theta}_i^{(t)} + \mathbf{v}_i^{(t+1)} $;
			
			\STATE Evaluate fitness value;\\
			
			\STATE Calculate the historical optimal position of particle $ i $: \\
			$ \mathbf{p}_i^{(t+1)} = \left\{
			\begin{array}{lc}
				\mathbf{p}_i^{(t)}, & R'\big(\mathbf{p}_i^{(t)}\big) \le R'\big( \bm{\theta}_i^{(t+1)} \big)\\
				\bm{\theta}_i^{(t+1)}, &  R'\big(\mathbf{p}_i^{(t)}\big) > R'\big( \bm{\theta}_i^{(t+1)} \big)	
			\end{array}
			\right.$;	
			
			\STATE Find $R'( \mathbf{p}^{*(t+1)}  ) \! = \! \min\{ \!R'( \mathbf{p}_1^{(t+1)}),\!\dots\!,R'( \mathbf{p}_L^{(t+1)}) \! \}    $;\\
			
			\ENDFOR
			
			\STATE Calculate the historical optimal position of the population:\\
			$ \mathbf{g}^{(t+1)} = \left\{
			\begin{array}{lc}
				\mathbf{g}^{(t)}, & R'\left(\mathbf{g}^{(t)}\right) \le R'\big( \mathbf{p}^{*(t+1)} \big)\\
				\mathbf{p}^{*(t+1)}, &  R'\left(\mathbf{g}^{(t)}\right) > R'\big( \mathbf{p}^{*(t+1)} \big)
			\end{array}
			\right. $;\\
			
			\STATE Adjust adaptive parameter shown in Algorithm \ref{alg:AAP};\\
			\STATE Set $ t \gets t+1 $.
			\ENDWHILE
		\end{algorithmic}
	\end{algorithm}

The fitness value of each particle is evaluated by using the fitness function $ R'({\bm{\theta}}) $, which can be defined as follows:
\vspace{-0.8ex}
$$ R'({\bm{\theta}}) = -R_{\rm{sum}}. \vspace{-0.8ex} $$
%inspired by the bird swarm motion model. 
%It regards all possible solutions in the solution space as the habitat of the bird swarm movement model, then gradually improves the possibility of finding better solutions in the solution process by interacting the information between individuals, and guides all particles in the population to gather towards the possible solutions.
%Supposing the size of particle population is $ L$, the decision space has dimension of $ N $  and the maximum number of iterations is $ T $.
%For $ i=1,2,\dots,L $, the coordinate position of particle $ i $ at time $ t $ can be expressed as $  {\bm{\theta }}_i^{(t)} = (\theta_{i1}, \theta_{i2}, \dots , \theta_{iN}  )  $. The fitness value of each particle is evaluated by using the fitness function $ R'({\bm{\theta}}) $, which can be defined as follows:
%\vspace{-0.8ex}
%$$ R'({\bm{\theta}}) = -R_{\rm{sum}}. \vspace{-0.8ex} $$
%For CPS, each elemrnt of $ {\bm{\theta }}_i^{(t)}  $ is limited within [0, $ 2\pi $] while for DPS, the phase shifts are descreted and then limited.
Finding the maximum value of $ R_{\rm{sum}} $  means finding the minimum value of $ R'({\bm{\theta}}) $, then the minimum value of the reciprocal of the exponential product will be found accordingly.
%and limit the position boundary of each element in $ {\bm{\theta }}_i^{(t)} $ within $ [0,2\pi] $.

The velocity of particle $ i $ is defined as the distance of particles moving in each iteration, expressed as
$  \mathbf{v}_i^{(t)} = (v_{i1}, v_{i2}, \dots , v_{iN}  ) $, each of which is limited within $ [ -v_{\rm{max}},v_{\rm{max}}] $.
$ \mathbf{p}_i^{(t)} $ and $ \mathbf{g}^{(t)}  $ are respectively defined as the optimal position of particle $ i $ and the optimal position of the whole population after $ t $ iterations.
%$ Q $, $ Q_min=\max\{2, \lfloor p \rfloor \}  $, respectively represent the number of particles in the neighborhood and the minimum number among $Q$.
$ \omega  $  represents the inertia weight, which is used to adjust the search scope of the solution space and balance the global convergence and convergence rate. $ c $,  $ c_1 $ and $ c_2 $, and $ r_1 $ and $ r_2 $ respectively represent stagnation counter, acceleration constants, and random values within $ [0,1] $.

\vspace{-0.3cm}
{\setlength{\abovecaptionskip}{-0cm} 
	\setlength{\belowcaptionskip}{-0.2cm}  
	\begin{algorithm}[h]
		\caption{Adjust Adaptive Parameter}
		\label{alg:AAP}
		\setstretch{1}
		\vspace{-3ex} 
		\begin{multicols}{2}
			\begin{algorithmic}[1]
				
				\STATE 
				Initialize $ c=0 $;\\
				
				\IF { $ \mathbf{g}^{(t+1)} < \mathbf{g}^{(t)}$  }
				\STATE	$ flag=1 $;\\
				\ELSE
				\STATE $ flag=0 $; \\
				\ENDIF
				
				\IF {$ flag=0 $}
				\STATE $ c=c+1 $;
				\ELSE
				\STATE $ c=\max\{c-1,0\} $; 
				\IF {$ c < 2 $}
				\STATE $ \omega = 2\omega $;
				\ELSIF {$ c > 5 $}
				\STATE $ \omega = \omega/2 $;
				
				\ENDIF
				\ENDIF		
			\end{algorithmic}
		\end{multicols}
		\vspace{-2ex} 
	\end{algorithm}
}
%1) \textbf{Particle swarm initialization:}
%Initialize the population size and set the stopping criterion to reach the maximum number of iterations. The initial position and velocity of each particle are vectors composed of random numbers uniformly distributed within the allowable range. The historical optimal position of each particle is set as the initial position and the global optimal position is set as the best one among the individual optimal positions of all particles.

%2) \textbf{Evaluate particle fitness:} 

%3) \textbf{Calculate the historical optimal position of an individual:}
%The positions of each particle are brought in and the fitness function values of which are continuously compared with that of the individual optimal position obtained at the last iteration to update new optimal position of each particle.

%4) \textbf{Calculate the historical optimal position of population:}
%The optimal positions of each particle are brought in and the fitness function values of which are continuously compared with that of the global optimal position obtained at the last iteration to update new optimal position of the whole population.

%5) \textbf{Update the velocity and position of particles:}
%Particle swarm constantly adjusts and updates the velocity and position of the next step according to the historical optimal positions of individuals and groups.

\vspace{-0.2cm}
\section{Simulation Results}\label{simlresult}
In this section, we evaluate the impact of various parameters on the sum achievable rate performance. Our simulation parameters are set with reference to  \cite{panintelligent},\cite{zhi2021power}.
We assume the BS and the RIS are placed at $ (0,0) $ and $ (5,2) $ in a rectangular coordinate system, respectively. The users are uniformly and randomly scattered in a circle centered at $ (400, 0) $ with
radius of 4 \rm m. The AoD of users are  randomly generated from $ [0, 2\pi) $  and these angles will be fixed after initial generation.
The large-scale path loss model is modeled in dB as \cite{panintelligent}:
 {\setlength\abovedisplayskip{3pt}	
	\setlength\belowdisplayskip{3pt}
\begin{equation}
	PL = PL_0 \left( \frac{D}{D_0}   \right)^{-\kappa},
\end{equation}}
where $ PL_0 $ is the path loss at the reference distance $ D_0 $, $ D $ is the link length in meters, and $ \kappa $ is the path loss exponent. Here, we set the model parmeters as \cite{zhi2021power}: $ D_0=1 $, $ PL_0 = -30$ \rm dB, the path loss exponents of the BS-to-RIS and RIS-to-$ k $-th-user links are $ \kappa_{BI}= \kappa_{IU_k} = 2.8, \forall k $.
Unless otherwise stated, our simulation parameters are set as follows: number of users of $ K=6 $, number of antennas at the BS of $ M=64 $, 
number of reflecting elements of the  RIS of $ N=16 $,
transmit power of $ P=30 $ \rm dBm, noise power of $ \sigma^2 =-104 $ \rm dBm, Rician factor of $ K_G=1 $, $K_k = 10, \forall k $.
We also set $ d = \frac{\lambda}{2} $ in order to mitigate the spatial correlation between antennas.
The main parameters for PSO are: $ L=\min\{100,10N \}  $, $ T=200N $, $ v_{\rm{max}}=2\pi $, $ \omega=0.9 $,  $ c_1=c_2=1.49 $.

{\setlength{\abovecaptionskip}{-0.2cm}
	\setlength{\belowcaptionskip}{-0.5cm}  
	\begin{figure}
		\vspace{-0.7cm}
		\centering
		\includegraphics[scale=0.45]{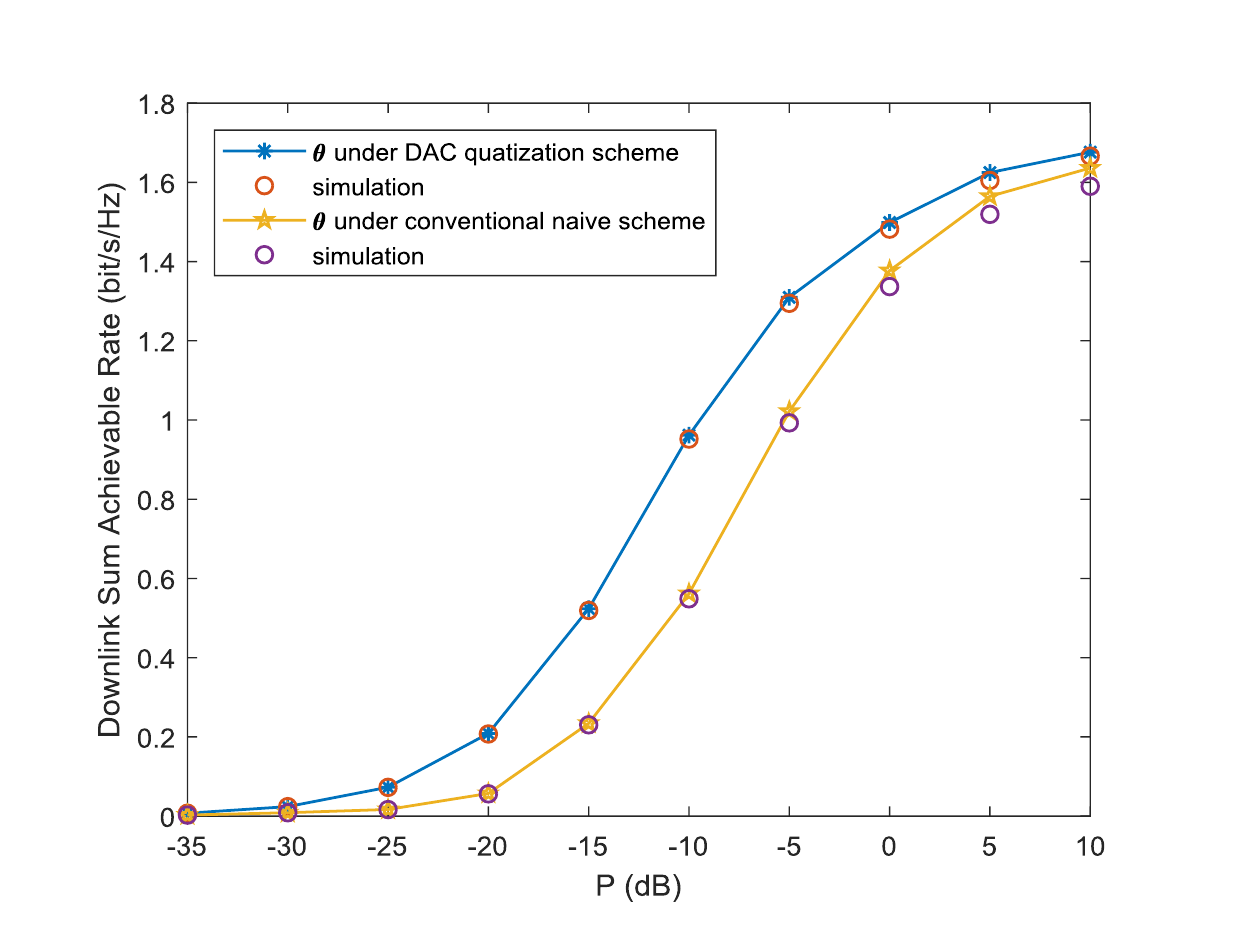}
		\caption{The downlink sum achievable rate versus $ P $ with $ b=1 $ under PSO.}
		\label{fig:P-R}
		\vspace{-0.2cm}
\end{figure}}

It is observed from Fig.~\ref{fig:P-R} that the derived results are consistent with the Monte-Carlo simulation results, which verify the correctness of the derived results. Specifically, we illustrate the downlink sum achievable rates versus transmit power, where one of the curves considers the hardware imbalance, while the other does not and naively regards the actual hardware imbalance as perfect. Specifically, for the conventional naive scheme, we first obtain the beamforming solution under the perfect hardware case, and then substitute the obtained solution into the SINR expression with actual hardware impairment. It is observed from this figure that the proposed algorithm is robust to the hardware impairment. 

Fig.~\ref{fig:b-R} shows the downlink sum achievable rate versus the resolution of DACs b. As shown in this figure, the achievable rates increase with b in both cases of CPSs and DPSs. The larger the quantization error, the lower the data rate. Moreover, the rates gradually converge to a constant, which is the achievable rate obtained in the case of $ b \to \infty $. It shows that four quantization bits are enough to capture a large portion of the performance of the ideal perfect DACs case. 
%We also observe that the larger the $ N $, the smoother the curve trend, which means the lower error caused by low-resolution DACs.
{   
	\setlength{\abovecaptionskip}{-0.2cm}
	\setlength{\belowcaptionskip}{-0.5cm} 
	\begin{figure} 
		\vspace{-0.7cm}
		\centering
		\includegraphics[scale=0.45]{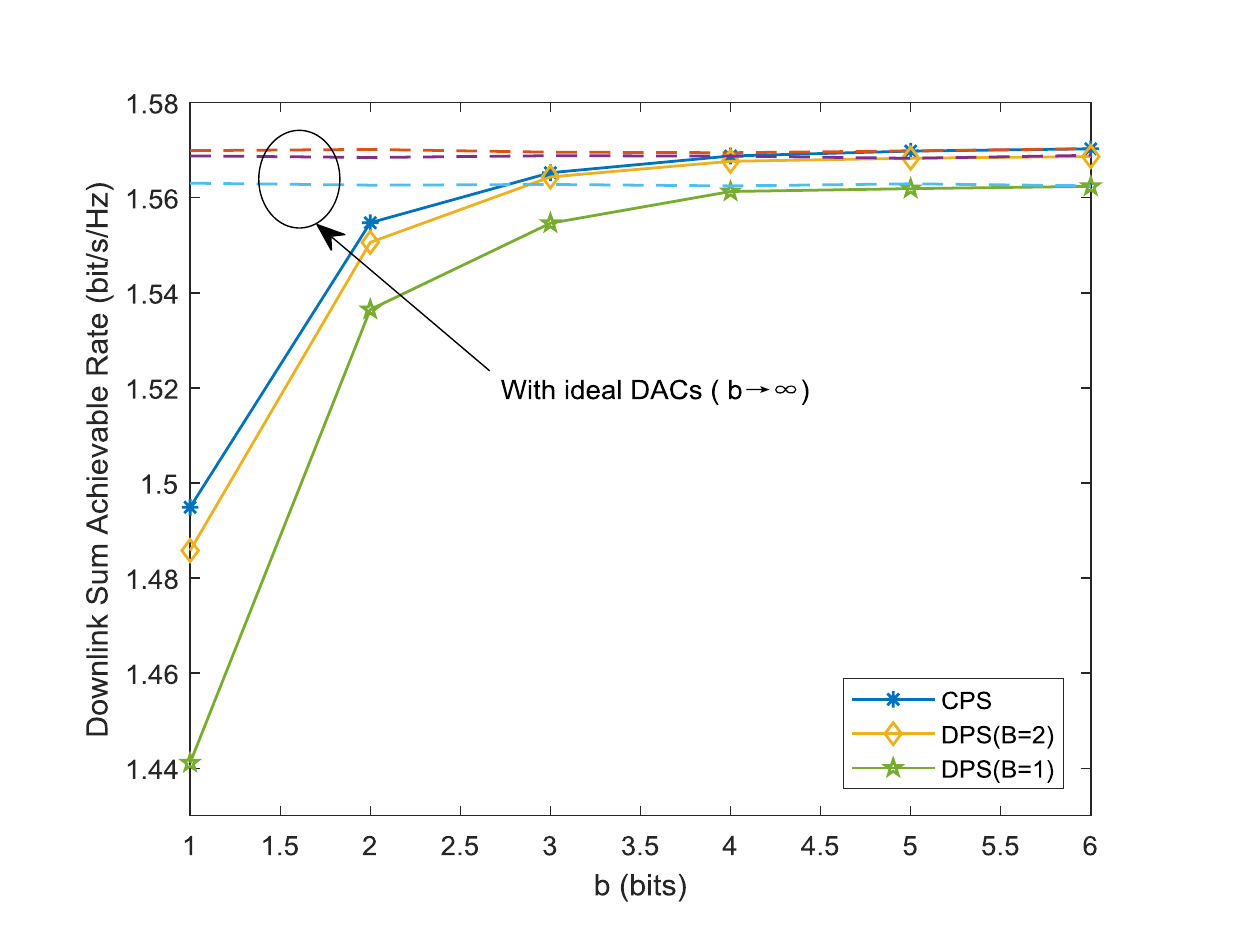}
		\caption{The downlink sum achievable rate versus the number of quantization bits of DACs.}
		\label{fig:b-R}
		\vspace{-0.1cm}
\end{figure}}

{\setlength{\abovecaptionskip}{-0.2cm} 
	\setlength{\belowcaptionskip}{-0.5cm} 
\begin{figure} 
	\vspace{-0.6cm}
	\centering
	\includegraphics[scale=0.45]{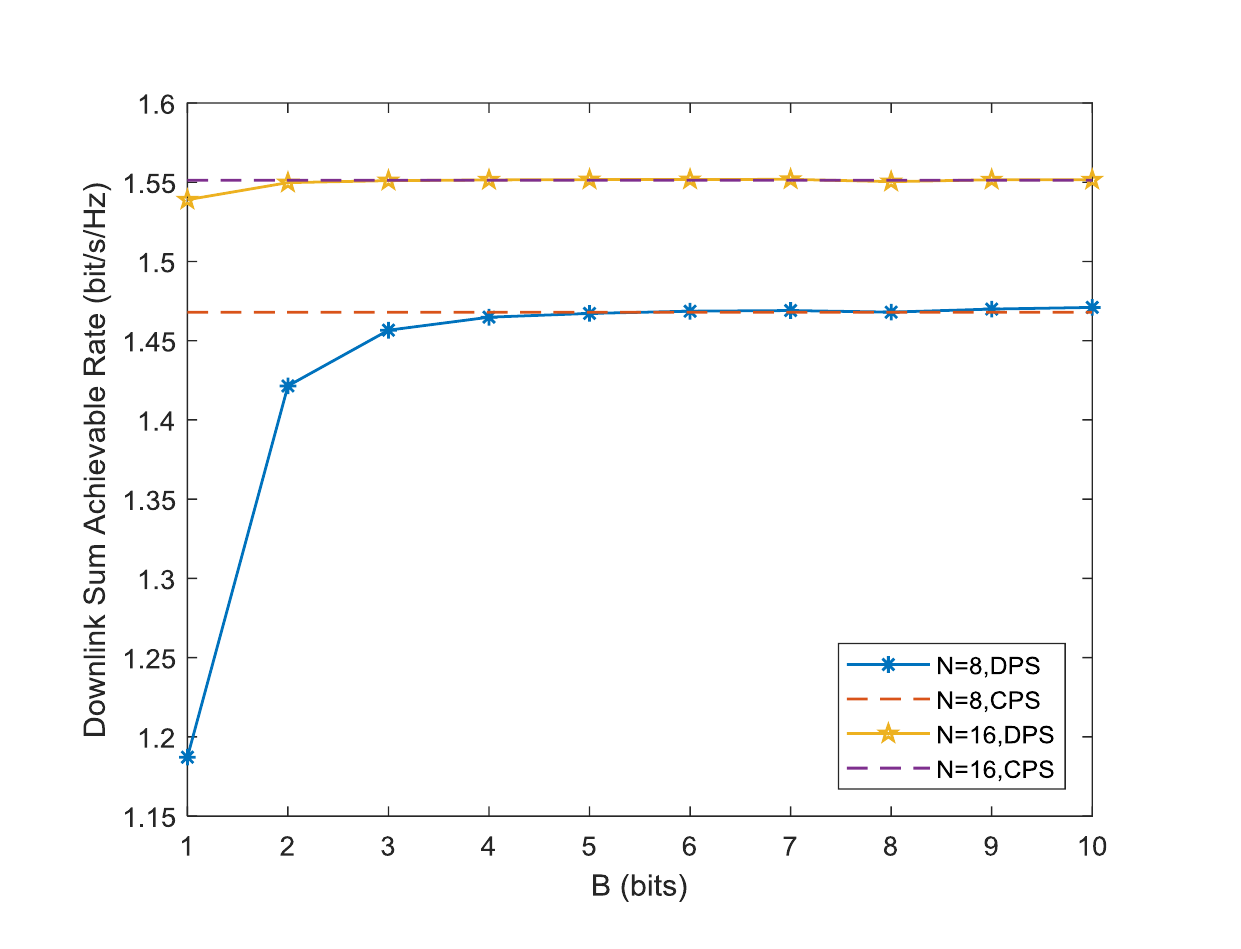}
	\caption{The downlink sum achievable rate versus the number of quantization bits of phase shifts at the RIS.}
	\label{fig:RISB-R}
		\vspace{-0.1cm}
\end{figure}}

In fig.~\ref{fig:RISB-R}, we fix $ b=1 $ and compare the sum achievable rate with the number of quantization bits of phase shifts at the RIS under different $ N $. The sum rate increases rapidly when $ B $ is small, while the curve gradually saturates when $ B $ becomes larger. In addition, when $ N $ is large, $ B $ has a marginal impact on sum rate.

%\begin{figure} 
%	\centering
%	\includegraphics[scale=0.45]{./fig./M-R.pdf}
%	\caption{The sum achievable rate of downlink versus the number of BS antennas.}
%	\label{fig:M-R}
%\end{figure}
%Fig.~\ref{fig:M-R} illustrate that the sum achievable rates increase with the number of BS antennas $ M $ and the upward trend is relatively flat when $ N $ is large. In addition, for three different simulation results of quantization bit of DACs $ b $ with 1, 2 and $ \infty $, the gap between curves decreases with the increase of $ b $, which means that the performance improved by increasing quantization bit $ b $ is very limited.

\vspace{-0.2cm}\section{Conclusion}\label{conclusion}
In this paper, a multi-user  massive MIMO system aided by a RIS has been discussed, in which each transmit antenna of the BS is equipped with a DAC. 
The simulation results have proved the correctness of the derived achievable rate and the superiority of using the algorithm when considering low-resolution DACs.
A DAC with almost four bits is sufficient to achieve the same rate as an ideal DAC, which verify the rationality of using low-resolution DAC in the system.
%Based on Rician fading channel, the approximate sum achievable rate has been derived based on MRT and AQNM. The PSO has been applied to optimize the phase shifts of the RIS under CPSs and DPSs to obtain the maximum sum achievable rate. 
%Finally, the influence of the quantization bits of the DACs and phase shifts at the RIS on the rate performance have been analyzed by simulation. The simultion results have shown that four quantization bits of DACs are enough to capture a large portion of the performance of the ideal perfect DAC case. Therefore, it is feasible to apply low-resolution DACs to the RIS-aided massive MIMO systems.
\vspace{-0.32cm}
\bibliographystyle{ieeetr}
% argument is your BibTeX string definitions and bibliography database\left(s\right)
\bibliography{refer}

%\section*{Acknowledgment}
%This work is supported by National 863 High Technology Development Project \left(No. 2013AA013601\right), Key Special Project of National Science and Technology \left(No. 2013ZX03003006\right) and National Nature Science Foundation of China \left(Nos. 61172077 \& 61223001\right).

%\numberwithin{equation}{section}
%\begin{appendices}
%\section{Proof of Theorem 3}

%\end{appendices}

\
\




\end{document}